\documentclass[a4paper,10pt,twocolumn]{article}     

\usepackage{graphicx}
\usepackage{amsmath}
\usepackage{amssymb}
\usepackage{url}
\usepackage[round,authoryear]{natbib}

\newcommand{\bs}[1]{\mathbf{#1}}
\newcommand{\tr}{T}
\newcommand{\eq}[1]{\begin{equation} #1 \end{equation}}

\newcommand{\R}{\mathbb{R}}
\newcommand{\N}{\mathcal{N}}

\newcommand{\pe}{\varepsilon}
\newcommand{\peg}{\tilde{\pe}}
\newcommand{\sh}{x}
\newcommand{\so}{y}
\newcommand{\sog}{\tilde{\so}}
\newcommand{\vo}{\bs{y}}
\newcommand{\vh}{\bs{x}}
\newcommand{\vog}{\tilde{\vo}}
\newcommand{\vhg}{\tilde{\vh}}
\newcommand{\vzg}{\tilde{\bs{z}}}
\newcommand{\Mo}{\bs{Y}}

\newcommand{\vect}{\mathrm{vec}}

\begin{document}

\title{Recognizing recurrent neural networks (rRNN):\\Bayesian inference for recurrent neural networks
}

\author{Sebastian Bitzer \qquad Stefan J. Kiebel\\\\
        MPI for Human Cognitive and Brain Sciences \\
              Stephanstr. 1a, 04107 Leipzig\\
              Tel.: +49-341-99402426, Fax: +49-341-99402221\\
              \texttt{bitzer@cbs.mpg.de}, \texttt{kiebel@cbs.mpg.de}
}

\date{June 28, 2011}

\maketitle

\begin{abstract}
Recurrent neural networks (RNNs) are widely used in computational neuroscience and machine learning applications. In an RNN, each neuron computes its output as a nonlinear function of its integrated input. While the importance of RNNs, especially as models of brain processing, is undisputed, it is also widely acknowledged that the computations in standard RNN models may be an over-simplification of what real neuronal networks compute. Here, we suggest that the RNN approach may be made both neurobiologically more plausible and computationally more powerful by its fusion with Bayesian inference techniques for nonlinear dynamical systems. In this scheme, we use an RNN as a generative model of dynamic input caused by the environment, e.g. of speech or kinematics. Given this generative RNN model, we derive Bayesian update equations that can decode its output. Critically, these updates define a 'recognizing RNN' (rRNN), in which neurons compute and exchange prediction and prediction error messages. The rRNN has several desirable features that a conventional RNN does not have, for example, fast decoding of dynamic stimuli and robustness to initial conditions and noise. Furthermore, it implements a predictive coding scheme for dynamic inputs. We suggest that the Bayesian inversion of recurrent neural networks may be useful both as a model of brain function and as a machine learning tool. We illustrate the use of the rRNN by an application to the online decoding (i.e. recognition) of human kinematics.
\end{abstract}
\section{Introduction}
Recurrent neural networks (RNN) have been used for many years now to augment nonlinear mappings with a dynamic representation \citep{Pearlmutter1989,Williams1989,Narendra1990,Jaeger2001,Maass2002}, for example, for the classification of sensory input in machine learning. In computational neuroscience, RNNs are extensively used to investigate the dynamic properties of cortical networks \citep[e.g.][]{Buonomano2009,Legenstein2007}, to model  the measured activity of networks of neurons \citep[e.g.][]{Friston2003,Kiebel2006,Kiebel2009c,Sotero2007,Rodrigues2010} and more generally to model brain processes like perception, memory and attention \citep[e.g.][]{Elman1990,Miller2001,Hamker2005}. The recurrent connections of these networks capture two of the most prominent features of neuronal networks observed in the brain: Firstly, connections between two neurons are rarely uni-directional but more often bi-directional, potentially via more than one synapse. Secondly, neurons perform highly nonlinear operations, i.e. they transform their input to spiking output. Recurrent neural networks capture both these features where often the input (post-synaptic potentials) and output (action potentials) are replaced by summary measures, i.e. the post-membrane potential function and firing rate. In such a continuous-time RNN, each neuron (often called unit) performs a simple operation: In each moment in time, it applies a nonlinear function to the sum of its input and passes this on to other neurons. This simple mechanism can provide for extremely rich patterns of activity in each neuron, even with a network of small size. Literally thousands of contributions in computational neuroscience and machine learning are based on networks of these firing rate-coding units \citep{Rabinovich2006,Cessac2007}. 

As powerful as RNNs are as a model class, recent contributions have questioned the usefulness of RNNs to explain neurobiological phenomena observed in real networks of neurons \citep{Spruston2008,Mel2008,Debanne2011}. Here, we suggest that a simple re-interpretation of the functional role of RNN dynamics leads to a novel and potentially more plausible account of what RNN units may compute: We suggest that neuronal networks serve as Bayesian decoders of dynamics caused by the environment. For example, in action observation, humans decode the kinematics of other people from visual input dynamics. For Bayesian recognition, one needs a so-called generative model for the hidden dynamics of the environment which cause sensory input to the brain. We suggest that RNNs are an ideal generative model for these hidden dynamics in our environment. The task of the recognition system is to decode the sensory input generated by the hidden RNN dynamics. To do this, we derive Bayesian update equations from the generative RNN model and call these 'recognizing RNN'. The difference to a standard RNN is that each unit in the recognizing RNN computes more sophisticated updates involving predictions and prediction error messages from other units in the network. Here we show that a recognizing RNN can decode real-world dynamics (human kinematics) and display several features which can also be observed with real neuronal systems, e.g., the online decoding of hidden dynamics in the environment, computation of predictions and prediction error, robustness to noisy input and fast adaptation to sudden changes in the environment. These features are not only general hallmarks of brain function but may, in principle, also be useful for machine learning applications for decoding dynamics in an online fashion.

In computational neuroscience, models of recurrently connected networks of neurons, which optimally estimate dynamically changing states from noisy observations, have recently been proposed \citep{Rao2004, Deneve2007, Natarajan2008, Wilson2009, Boerlin2011}. While these models provide important insights, results were reported for relatively restrictive conditions such as linear dynamics \citep{Deneve2007,Wilson2009,Boerlin2011}, discrete states \citep{Rao2004,Deneve2007,Boerlin2011}, or a one-di\-men\-sion\-al state \citep{Natarajan2008,Wilson2009,Boerlin2011}. Although \citet{Natarajan2008} allow for nonlinear dynamics they assume knowledge of an ideal observer which provides an instantaneous error signal for learning of network connections. Similarly, reservoir computing approaches \citep{Jaeger2001,Maass2002,Verstraeten2007} rely on a teaching signal which provides a desired output at every point in time during learning. In contrast, we propose an approach combining multi-dimensional, continuous-time hidden nonlinear dynamics where learning proceeds without an externally provided error signal. Our main contribution is to demonstrate that a recognizing RNN is well suited to recognize dynamic stimuli and may be used as a functional model for neuronal ensemble dynamics. In particular, we will illustrate this by showing that the prediction errors computed by a recognizing RNN provide sufficient information to discriminate dynamic stimuli, in an online fashion. 

The present approach may also lead to a better understanding of the role of recurrently connected networks of neurons in the brain: predictive coding has been suggested as a theory for hierarchical processing in the brain in which different levels exchange prediction and prediction error messages \citep{Mumford1996,Rao1997,Rao1999,Friston2009}. \citet{Rao1997} already described RNN-like dynamic models to implement predictive coding for static stimuli. The present approach can be seen as an extension to Rao and Ballard's original work to provide inference for \emph{dynamic} stimuli by resorting to approximate inference methods for nonlinear, continuous dynamic models \citep{Friston2008a}.

The remainder of the paper is organised as follows. In Materials and Methods we (i) present  RNNs as generative models, (ii) describe the Bayesian inference framework and (iii) show that dynamic updates of the posterior state critically depend on prediction error. We illustrate the rRNN approach using human motion capture data. In Results we demonstrate that the rRNN can successfully recognize human kinematics and discriminate between different walking styles based on the prediction error of rRNN units.
\section{Materials and Methods}
In the following, we will describe the two key elements of the present approach: a recurrent neural network (RNN) as a generative model of the sensory dynamics and the Bayesian inference framework to derive the update equations for a recognizing RNN (rRNN). Subsequently, we will apply the rRNN technique to the recognition of human kinematics, for which we describe the kinematic data and the rRNN settings. 

To motivate the present approach, we will start with a brief summary of the conventional RNN technique as used in machine learning for classification of stimuli. Note, however, that it is not our aim to compare discrimination performance of conventional and recognizing RNNs. Rather, description of the conventional RNN is given as a reference for understanding the conceptual differences between the two approaches.
\begin{figure*}
  \begin{center}
    \includegraphics[width=\textwidth]{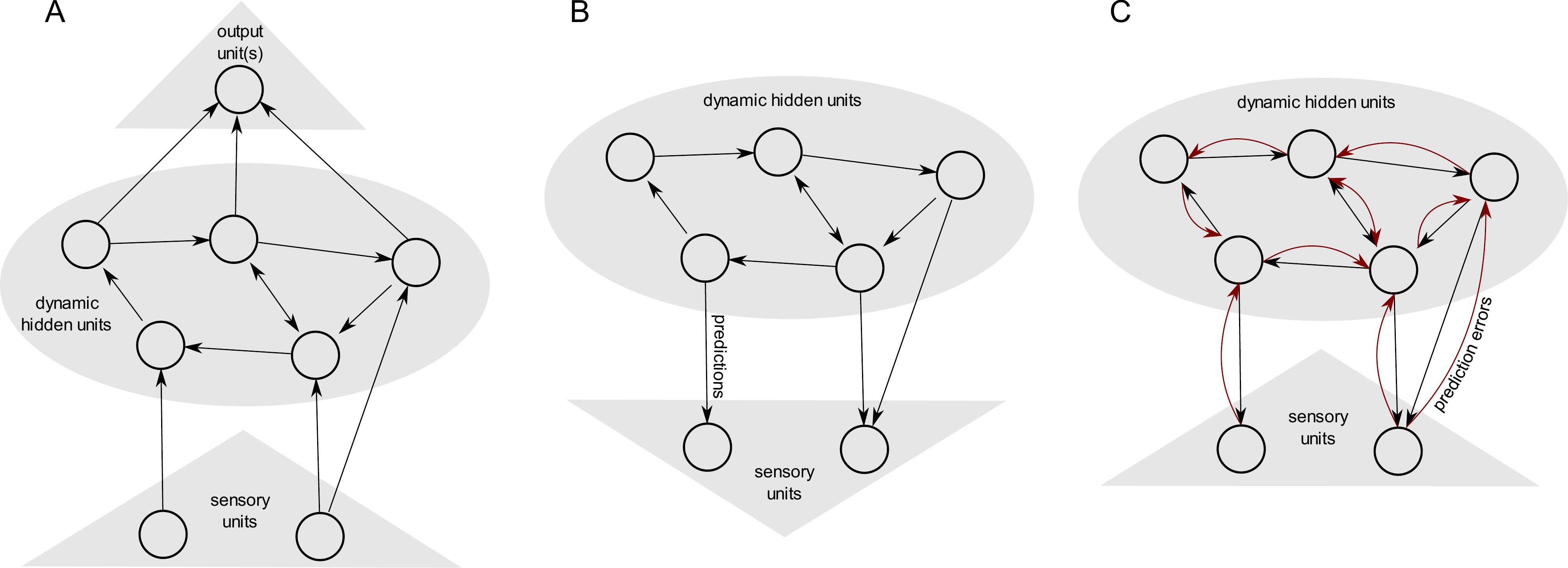}
  \end{center}
  \caption{Comparison of different RNN architectures. (A) conventional RNN, (B) generative RNN (gRNN) and (C) recognizing RNN (rRNN). Each RNN has dynamic hidden units, but the overall direction of information flow differs (indicated by the grey triangles). The conventional RNN is designed to compute an output given sensory input. In contrast, the gRNN computes sensory states. Finally, the rRNN computes predictions (black arrows) and prediction error messages (red arrows) to recognize the hidden causes that generated the sensory input. 
}
  \label{fig:rnns}
\end{figure*}
\subsection{Conventional Recurrent Neural Network}
The RNN technique has been used in many machine learning applications such as classification of static or dynamic stimuli, or time-series prediction. This approach has a long history which took off with the development of a supervised learning routine \citep{Pearlmutter1989,Williams1989}. Recently, this learning approach has been complemented by the so-called reservoir computing technique \citep{Jaeger2001,Maass2002}.

In general, in a conventional RNN, sensory units provide input, which drives the dynamics of the hidden units (see Fig. \ref{fig:rnns}A). Output units readout the result of the dynamic computations based on a mixing of the sensory and hidden states. 

An example of such a network is discussed in \citep{Jaeger2007} where the continuous-time dynamics based on leaky-integrator units is given by
\eq{\label{eq:RNNdyn}\begin{split}
    \dot{\sh}_i &= f(\vh)_i\\
    &= k_i\left(-a\sh_i + \tanh([\bs{W}^\mathrm{in}\vo + \bs{W}\vh + \bs{W}^\mathrm{fb}\bs{o}]_i) \right)
\end{split}}
where $\sh_i$ is the state of hidden unit $i \in \{1,\dots,H\}$, $\vo\in \R^{I\times 1}$ are the states of the input (sensory) units, $\bs{o} \in \R^{D\times 1}$ are the states of the readout units, $\bs{W} \in \R^{H\times H}$ is a weight matrix defining the interaction between the $H$ hidden units, similarly $\bs{W}^\mathrm{in}$ and $\bs{W}^\mathrm{fb}$ define the connections from the input to the hidden units and the (optional) feedback connections from the readout units, respectively. $k_i$ is a rate constant for unit $i$ and $a$ is the amount of leakage. The output states are determined by 
\eq{\label{eq:RNNout}
    \bs{o} = \bs{V}[\vh^\tr,\vo^\tr]^\tr
}
where $\bs{V} \in \R^{D\times (H+I)}$ is a weight matrix.

In a conventional RNN, the overall flow of information is from sensory to output units, because the RNN serves as a model for neuronal dynamics (hidden states) which are used to compute, e.g., a classification of the sensory input. We now use the same dynamics where we reverse the flow of information to model the generation of sensory dynamics by hidden states of the environment (e.g. body movements cause visual output dynamics). 
\subsection{Generative Recurrent Neural Network}
Our overall aim is to construct a recognition system which can recognize its sensory observations based on its internal dynamics. For a Bayesian recognition system we require a dynamic generative model, for which we choose a RNN. This 'generative recurrent neural network' (gRNN) runs independently of any input and generates sensory data, i.e. observations. Note that, in comparison to a conventional RNN (Eq. \ref{eq:RNNdyn}), here the sensory units become the output of the network while no input units are defined (hence the missing units which acted as output in the conventional RNN). Consequently, the flow of information is reversed in the gRNN and its autonomously running hidden dynamics drive its sensory units (see Fig. \ref{fig:rnns}B). In particular, we define a gRNN as
\eq{\label{eq:gRNNdyn}
    \dot{\sh}_i = f(\vh)_i = k_i\left(-a\sh_i + \tanh([\bs{W}\vh]_i) \right)
}
\eq{\label{eq:gRNNobs}
    \vo = \bs{V}\vh
}
where now $\bs{V} \in \R^{D\times H}$ linearly translates hidden states $\vh$ into sensory states $\vo$. This gRNN computes sensory output $\vo$ as caused by a hidden, dynamic process defined by the RNN dynamics $f(\vh)$. In the following section, we describe how a recognizing recurrent neural network (rRNN) is constructed from the gRNN using Bayesian inversion. This rRNN receives sensory observations (as in a conventional RNN, Eq. \ref{eq:RNNdyn}) and infers about the hidden states that caused these observations. Effectively, conventional RNN computations are aimed at doing the same (c.f. Fig. \ref{fig:rnns}A,C); however, the update equations of an rRNN are explicitly derived for this recognition task.
\subsection{Recognizing Recurrent Neural Network}
Generative models like a gRNN (eqs. \ref{eq:gRNNdyn},\ref{eq:gRNNobs})  can be used to derive a Bayesian inference system that recognizes input caused by the generative model. In the probabilistic setting, when observations and state transitions are noisy, or uncertain, Bayesian inversion of the generative model leads to updates of the hidden states which make them an optimal representation of the observations given the model. For example, the well-known Kalman-Bucy filter \citep{Jazwinski1970} implements such a Bayesian inversion scheme for linear dynamic processes. The gRNN uses highly nonlinear dynamics (Eq. \ref{eq:gRNNdyn}) and, therefore, we require approximate inversion schemes \citep{Jazwinski1970,Wan2001a,Doucet2001,Friston2008a,Daunizeau2009,Friston2010c}. Here, we derived the update equations using the D-step of Friston's dynamic expectation maximization (DEM) framework \citep{Friston2008a}. This choice was based on our previous experience with inversion of continuous-time dynamic models using DEM \citep{Kiebel2009a} but, in principle, other inversion schemes could be used as well. DEM uses generalised coordinates, local linearisation and point-estimates at strategically important positions. See the appendix for a high-level derivation of the algorithm and an explanation of generalised coordinates which are a dynamically extended representation of state variables, the use of which we indicate by a tilde in the subsequent formulas. 

In the following, we will briefly describe the key computations performed by DEM. This description is aimed at giving an intuitive description of the update equations governing the rRNN and will allow interpretation of these updates in terms of prediction and prediction error messages.

The most important equation resulting from inversion with DEM describes the evolution of the posterior mode of the hidden states in generalised coordinates and is given by
\eq{\label{eq:postmode}
    \dot{\vhg} = \kappa \frac{\partial V(\vhg)}{\partial \vhg} + \bs{D}\vhg.
}
The motion defined in this equation consists of two parts: 1) $\bs{D}\vhg$ which, in absence of other contributions,  implements that the motion of the posterior mode follows its local trajectory represented in generalised coordinates using a derivative operator $ \bs{D}$ and 2) the derivative of the variational energy $ V(\vhg)$ with respect to hidden states which acts as a corrective force to make the motion consistent with the gRNN and the observations. With fixed parameters, the variational energy is the log-joint probability of observations (sensory states) $\vog$ and hidden states $\vhg$ which defines the probabilistic gRNN. In particular, the variational energy is given by
\eq{\begin{split}
    V(\vhg) &= \log p(\vog,\vhg|\theta)\\
    &= \log p(\vog|\vhg,\theta) + \log p(\bs{D}\vhg|\tilde{\bs{f}},\theta)\\
    &= \log p(\vhg|\vog,\theta) + c
\end{split}
}
where $c$ is a constant, $\theta$ is a vector consisting of all parameters of the model and $\tilde{\bs{f}}$ are the dynamic predictions defined by Eq. \ref{eq:gRNNdyn} in generalised coordinates. The last term illustrates that the updates are a dynamic form of maximum a posteriori estimation of hidden states. Gaussian distributions are assumed for the state transition and observation densities:
\eq{
    p(\bs{D}\vhg|\tilde{\bs{f}},\theta) \sim \N\left(\tilde{\bs{f}}, \tilde{\Sigma}_\sh\right)
}
\eq{
    p(\vog|\vhg,\theta) \sim \N\left((\bs{I}\otimes \bs{V})\vhg, \tilde{\Sigma}_\so \right)
}
where $(\bs{I}\otimes \bs{V})\vhg $ is the predicted sensory state given the hidden states as defined by Eq. \ref{eq:gRNNobs} in generalised coordinates\footnote{$\bs{I}$ is the identity matrix with size equal to the number of used generalised coordinates and $\otimes$ is the Kronecker product}, and $\tilde{\Sigma}_\sh$ and $\tilde{\Sigma}_\so$ are the prior covariances of sensory and hidden states in generalised coordinates, respectively. This leads to a simple interpretation of the posterior mode updates in terms of prediction errors. In particular, the gradients of these densities with respect to hidden states become
\eq{\label{eq:Vgradh}\begin{split}
    \frac{\partial \log p(\bs{D}\vhg|\tilde{\bs{f}},\theta)}{\partial \vhg} &=  -\frac{1}{2}\frac{\partial}{\partial \vhg} \peg_\sh^\tr \tilde{\Sigma}_\sh^{-1} \peg_\sh\\
     &= -\left[\frac{\partial \peg_\sh}{\partial \vhg}\right]^\tr \tilde{\Sigma}_\sh^{-1} \peg_\sh
\end{split}}
\eq{\label{eq:Vgrado}\begin{split}
    \frac{\partial \log p(\vog|\vhg,\theta)}{\partial \vhg} &= -\frac{1}{2}\frac{\partial}{\partial \vhg} \peg_\so^\tr \tilde{\Sigma}_\so^{-1} \peg_\so\\
    & = -\left[\frac{\partial \peg_\so}{\partial \vhg}\right]^\tr \tilde{\Sigma}_\so^{-1} \peg_\so
\end{split}}
where the prediction errors are defined as
\eq{\label{eq:predErrs}\begin{split}
    \peg_\sh &= \bs{D}\vhg - \tilde{\bs{f}}\\
    \peg_\so &= \vog - (\bs{I}\otimes \bs{V})\vhg.
\end{split}}
This means that the updates of the posterior hidden states follow the gradient of the prediction error with step sizes determined by the prediction error itself weighted by the prior precisions. The contribution from the prediction error on the sensory states, $\peg_\so$, ensures that the sensory states are well explained by the hidden states while the contribution from the prediction error on the hidden states, $\peg_\sh$, ensures that the posterior dynamics of hidden states as encoded by the generalised coordinates is consistent with the learnt model dynamics. In particular, for the first generalised coordinate, the prediction error
\eq{\label{eq:dynPredErr}
    \pe_\sh = \dot{\vh} - f(\vh)
}
ensures that the posterior velocity corresponds to the learnt, noise free hidden unit dynamics as defined in Eq. \ref{eq:gRNNdyn}. Conversely, we will argue below that a consistently large prediction error $\pe_\sh$ provides evidence for an inconsistency between observed and learnt dynamics and can be used to discriminate among different dynamic stimuli.

The question remains how the system got to know a suitable gRNN which generates specific sensory dynamics. In our experiments we let the system learn its generative model by adapting connectivity parameters $\bs{W}, \bs{V}$ and rate constants $\bs{k}$ using an approach which was developed for the identification of dynamical (neural-mass) systems \citep{Friston2003, Kiebel2009c} and is based on maximum a-posteriori estimation of the parameters \citep{Friston2002,Friston2002a}. See the appendix for details.  Note that this initial learning step is not our main point in this paper; any learning approach that successfully learns hidden gRNN dynamics to represent a given dynamic stimulus could be used here \citep[e.g.][]{Wan2001,Roweis2001,Valpola2002,Doucet2003,Archambeau2008,Friston2008a,Daunizeau2009,Kantas2009,Lazar2009,Schoen2011}, but also standard RNN learning may be used, if the hidden state dynamics is assumed to be deterministic during learning.
\subsubsection{Message passing in the rRNN}
\label{sec:invRNN}
The updates defined by equations \ref{eq:postmode}, \ref{eq:Vgradh}, \ref{eq:Vgrado} and \ref{eq:predErrs} can be interpreted as network dynamics based on messages sent by sensory and hidden units. Algebraically, this can be seen by exemplarily inspecting the observation density update equation, Eq. \ref{eq:Vgrado}, for the first generalised coordinate of a single hidden unit $i$:
\eq{\label{eq:PEmessageObs}\begin{split}
    \frac{\partial \log p(\vog|\vhg,\theta)}{\partial \sh_i} &= -\frac{\partial \peg_\so^\tr}{\partial \sh_i} \tilde{\Sigma}_\so^{-1} \peg_\so\\
    & = -\sum_j \frac{\partial [\peg_\so]_j}{\partial \sh_i} \left[\tilde{\Sigma}_\so^{-1} \peg_\so\right]_j
\end{split}}
where the sum over $j$ runs over sensory units $\so_j$ in generalised coordinates. Note that the partial derivative of the prediction error of sensory unit $j$ with respect to the state of hidden unit $i$ describes how a change in the state of unit $i$ effects the prediction error of unit $j$. Therefore, the state update for hidden unit $i$ is a weighted sum of these prediction error gradients where each element of this sum corresponds to a 'prediction error message' from a single sensory unit $j$. To compute the prediction error message a sensory unit first has to compute a prediction. This is done using the forward equation (\ref{eq:gRNNobs}) of the gRNN which is a weighted sum of the hidden states $\vhg$ where the weights are determined by the connectivity of the gRNN. In the following we call the elements of this sum 'prediction messages' which are sent from a hidden unit $\sh_i$ to a sensory unit $\so_j$. In summary, the update equations define  a recognizing RNN (rRNN) where a hidden unit sends prediction messages to connected sensory and hidden units such that these can compute prediction error messages which are returned to the hidden unit to update its state (see also Fig. \ref{fig:rnns}C). The updates resulting from the dynamics density, Eq. \ref{eq:Vgradh}, follow the same logic, where the hidden unit $\sh_j$ takes the place of sensory unit $\so_j$. Each hidden unit, therefore, sends and receives two kinds of messages: prediction and prediction error messages.
\subsubsection{Induced connectivity of the rRNN}
The connectivity matrices $W$ and $V$ of the gRNN (Eq. \ref{eq:gRNNdyn}, \ref{eq:gRNNobs}) are not necessarily the same as in the rRNN. Generally, the rRNN will have all connections of the gRNN plus the corresponding reciprocal connections, plus some additional ones. To see this, note that the prediction error messages in the rRNN in Eq. \ref{eq:PEmessageObs} are 0, when hidden unit $i$ is not connected to sensory unit $j$, i.e., when hidden unit $i$ has no direct influence on the computation of predictions in sensory unit $j$ (then $\frac{\partial [\peg_\so]_j}{\partial \sh_i}=0$). Only sensory units which receive a connection from a hidden unit $i$ in the gRNN will contribute messages containing the \emph{derivative} of the prediction error. However, in the rRNN, sensory units $j$, which are not connected in the gRNN to a hidden unit $i$, may also contribute messages, containing only their prediction error, through the weights computation  $w_j = [\tilde{\Sigma}_\so^{-1} \peg_\so]_j$. In particular, if the $j$-th row of the sensory precision matrix, $\tilde{\Sigma}_\so^{-1}$, has nonzero entries in positions other than $j$, e.g., $k$, the weight of sensory unit $j$ in the update equation (Eq. \ref{eq:PEmessageObs}) depends on the prediction error of unit $k$. In this case, sensory unit $k$ contributes to the update of hidden unit $i$, even though hidden unit $i$ is not connected to sensory unit $k$ in the gRNN. This means, that there is an additional connection from sensory unit $k$ to hidden unit $i$ in the rRNN. 

In conclusion, only if the covariance matrix $\tilde{\Sigma}_\so$ is diagonal, the connectivity matrix of sensory to hidden units in the rRNN will only contain those connections which are reciprocal to the hidden to sensory unit connections in the gRNN. Conversely, if there are off-diagonal entries in $\tilde{\Sigma}_\so$, there will be corresponding additional connections from sensory to hidden units in the rRNN, relative to the gRNN. The same considerations apply to the connectivity between hidden units. In summary, the connectivity of the rRNN directly follows from the gRNN, only if the units' states in the gRNN are a priori independent. For simplicity, this case is shown in Fig. \ref{fig:rnns}C and used in the following simulations. Note that a diagonal covariance matrix $\tilde{\Sigma}_\so$ is a natural assumption for the present data because we assume that the measurement noise is white and any correlation among observations is caused by the underlying dynamics which are modelled by the RNN dynamics.
\subsection{Human Movement Data}
\label{sec:walks}
\begin{figure*}
  \begin{center}
    \includegraphics[width=.6\textwidth]{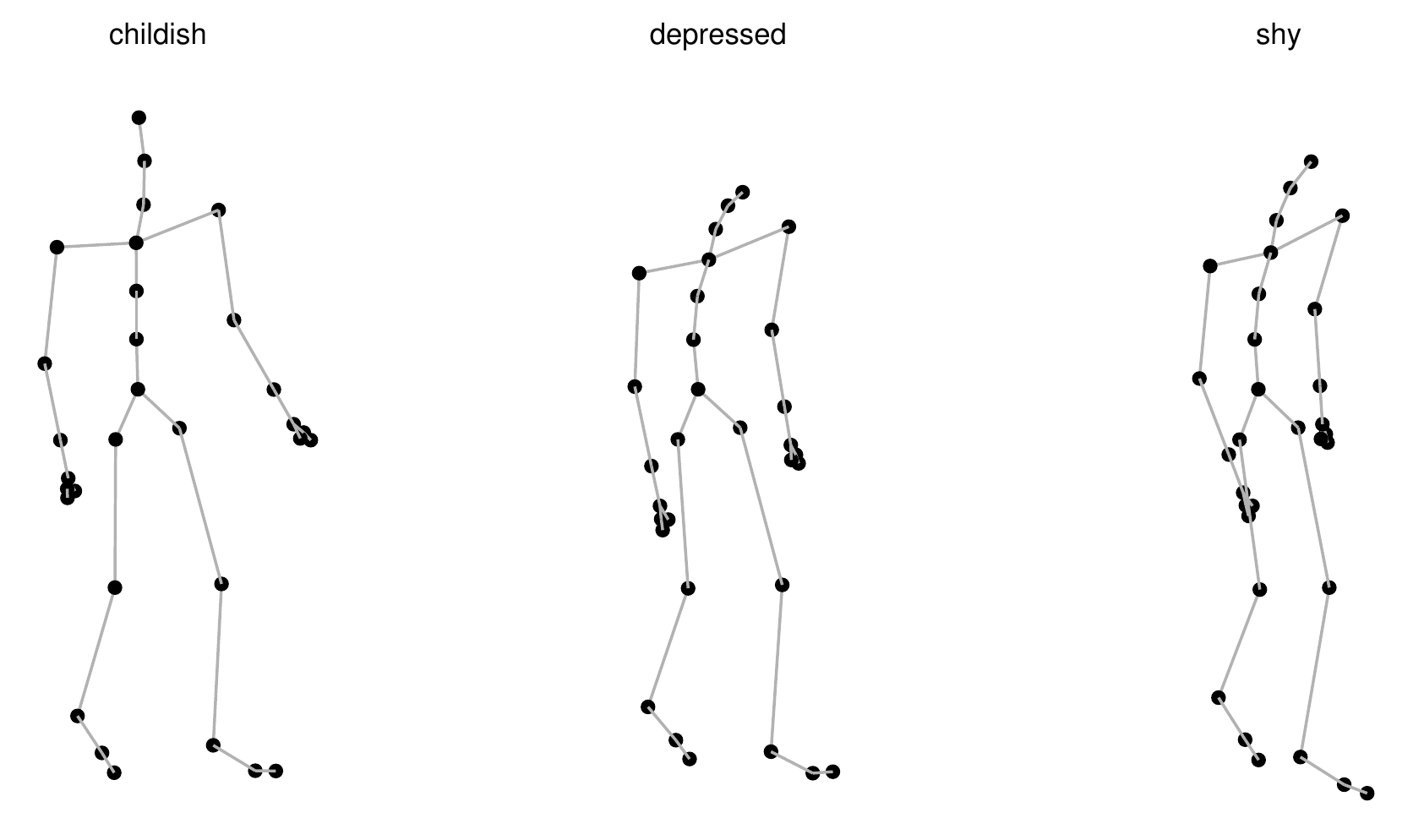}
  \end{center}
  \caption{
    Example frames from the three different walking styles: 'childish', 'depressed' and 'shy' (left to right). In our experiments we used the first five principal component coordinates of the motion capture 3D joint coordinates (indicated as filled circles) as observation variables. Lines are plotted only for visualisation purposes.
  }
  \label{fig:walkSkels}
\end{figure*}

We use human movements to demonstrate the properties of the rRNN in the experiments below. The kinematics of humans is highly dynamic and nonlinear through complicated interactions between individual joints. Kinematics, therefore, provides a good example of the kind of complex, dynamically changing, real-world stimuli which can be modelled using rRNNs. Here, we used three walks of the same subject, each of which expresses a different walking style (categorized as 'childish', 'depressed' and 'shy'; freely available from the CMU motion capture database, http://mocap.cs.cmu.edu, subject 142, motions 1, 5 and 19). We chose this particular subject because a large range of different movements were available among which we chose the selected walks because of their similar time-scales. The advantage of using motion capture data as compared to video is that we can focus on modelling the kinematics of the subject in terms of changing joint angles without the need to model detailed processing of visual information. 

For each walk we removed the global translation of the body and computed the 3D positions of the joints and extremities for all time points. This removed potential 'jumps' introduced by the circularity of the joint angles. As a result we obtained a set of 30 points moving in 3D space (see Fig. \ref{fig:walkSkels} for an example). Subsequently, we selected four representative seconds of data starting when the left foot touched the ground for each walk and subsampled the data using 30 frames per second resulting in $N=120$ data points per walk. These data covered roughly two footsteps for each movement. We then found a common, low-dimensional representation for the three walks using principal component analysis (of all three walks combined) which reduced the dimensionality from 90 dimensions to $D=5$ (maintaining $95.5\%$ of the original variance). Additionally, we scaled the coordinates of each walk such that the maximum absolute value in each dimension was 1 over all walks. In summary, we obtained for each of the three walks a sequential data set containing five trajectories each consisting of 120 time points, see Fig. \ref{fig:learnResult}.
\subsubsection{Learning of generative RNNs}
For each of the three walks, we constructed one gRNN by learning suitable parameters $\bs{W}, \bs{V}$ and $\bs{k}$ (eqs. \ref{eq:gRNNdyn} and \ref{eq:gRNNobs}) so that the dynamics of each generative RNN replicated the movement data. Each RNN had five sensory units, each of which generated one of the scaled principal component coordinates. In initial tests, we found that a network with $H=12$ hidden units was the smallest network which gave consistently acceptable learning results and we consequently used this network size in our experiments. These tests also showed that good learning results were obtained, if the hidden units were sparsely connected. In particular, we fixed 2/3 of all connections in $\bs{W}$ and 1/3 of all connections in $\bs{V}$ to 0. Other entries in $\bs{W}$, the rate constants $\bs{k}$ and the initial vector of hidden unit states $\vh(0)^l$ were chosen randomly before learning while $\bs{V}$ was initially chosen to correctly predict the first data point of a walk given $\vh(0)^l$. For details of this  initialisation and the learning procedure see the appendix. Note that any learning procedure could have been used here. The main point made by this initial learning step is that a dynamic representation for each walk can be found using RNNs with few units. 
\begin{figure*}
  \begin{center}
    \includegraphics[width=\textwidth]{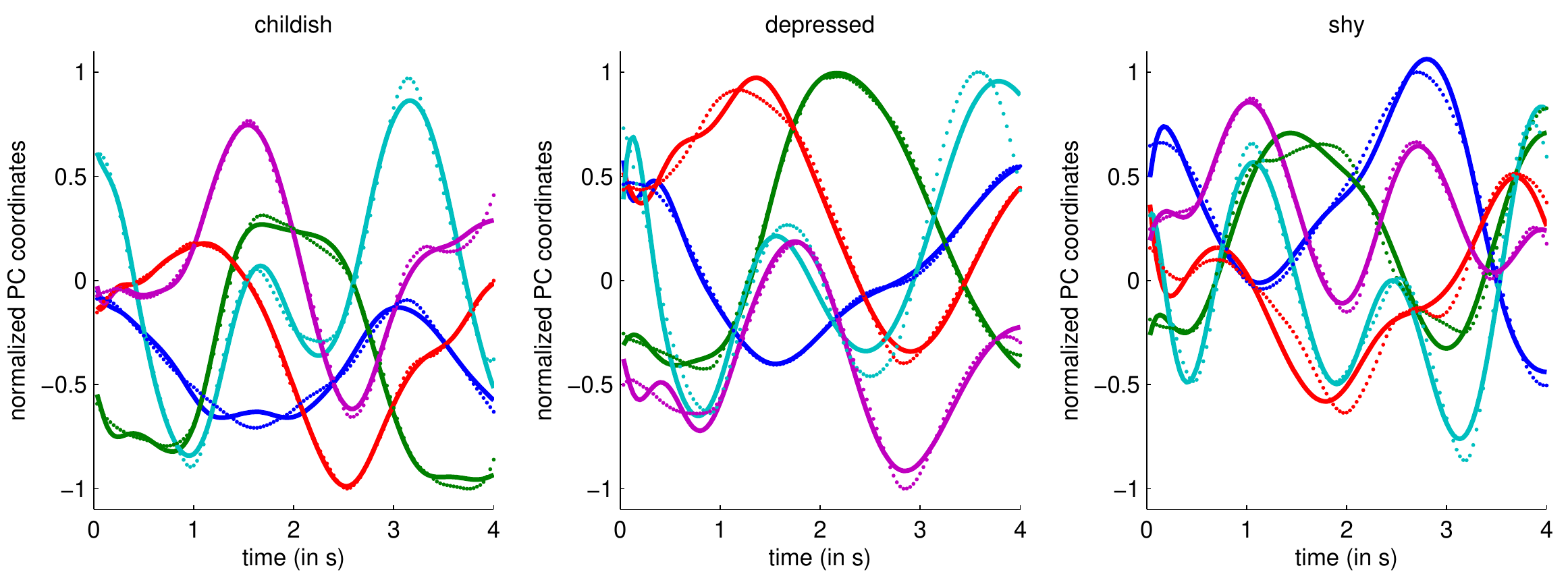}
  \end{center}
  \caption{
   Dynamics of three different human walks and model fits in principle component space (five components). Dotted lines: Original dynamics, Solid lines: Trajectories generated by a gRNN after learning. While the fit between data and its gRNN replications was not perfect, it was sufficiently close  such that the gRNN was an appropriate generative model for recognition (see Fig. \ref{fig:recognitionExmp}).
  }
  \label{fig:learnResult}
\end{figure*}

The sensory state trajectories of the learnt gRNNs  are shown in Fig. \ref{fig:learnResult}. Each of the three different walks was learnt well: The amount of variance explained for each walk was 99\%, 97\% and 97\% for the childish, depressed and shy walks, respectively.
\section{Results}
Here we demonstrate the utility of Bayesian inference for RNNs for online recognition of dynamic stimuli. As a proof of principle we apply the approach to the multi-dimensional, nonlinear kinematics of a walking human. We will first show that recognizing RNNs quickly and successfully recognize the hidden dynamics, i.e. decode the type of movement. Then, we will demonstrate that the prediction errors of the hidden units can be used to discriminate the three different walks. Finally, we will show that the rRNNs are robust against noisy observations and initial conditions. Note that all of the following experiments with the rRNN use the original motion capture data as observations.
\subsection{Fast Recognition of Dynamics}
\begin{figure*}
  \begin{center}
    \includegraphics[width=\textwidth]{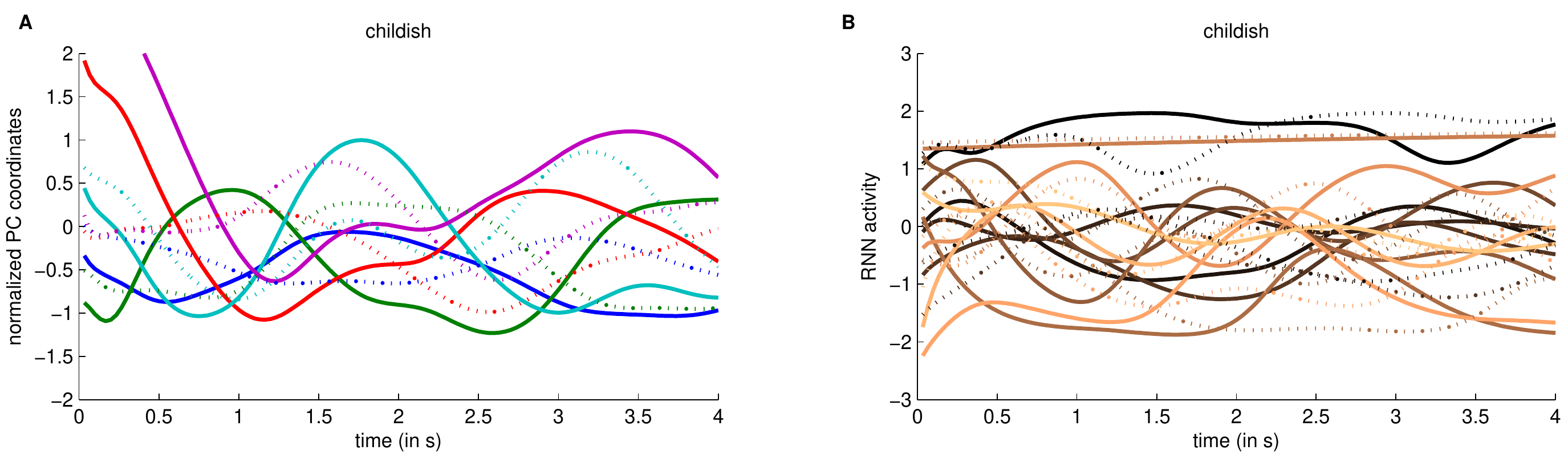}
  \end{center}
  \caption{
    Influence of initial hidden states on dynamics of generative RNN. Shown are trajectories of sensory (A) and hidden (B) states of the generative RNN for two different initial hidden states. Dotted trajectories resulted from initial hidden states used during learning ($\vh(0)^l$) while solid trajectories resulted from random initial hidden states ($\vh(0)^r$). In this example we used the RNN parameters learnt for walk 1 (childish), but results are qualitatively similar for other RNN parameters.
  }
  \label{fig:randTraj}
\end{figure*}

In this section, we show that the rRNN can quickly start recognizing a movement online. In particular, we show that this 'quick response' is robust against the initial hidden states at the beginning of the recognition process. This robustness is obtained despite the fact that gRNNs have a large dependence on their internal initial conditions. This is because RNNs are in general rich dynamic models which are capable of simultaneously representing many different dynamic stimuli depending on their initial conditions (hidden states). We demonstrate this for the gRNN for the childish walk. This gRNN was initialised during learning with the state $\vh(0)^l$. When this specific gRNN is started, after learning, in this state, the learnt shy walk is generated as shown in Fig. \ref{fig:learnResult} (left). However, when we initialise the same gRNN with a state $\vh(0)^r = \vh(0)^l + \epsilon$, which was perturbed by noise of the same size as the natural variability of the hidden states,  it generated very different trajectories of sensory states $\vo$ as well as hidden states $\vh$ as shown in Fig. \ref{fig:randTraj}. In other words, for deviating starting conditions, the gRNN generates dynamics that look different from the learnt kinematics and, when plotted in motion capture space, can deviate severely from a natural walk. 

\begin{figure*}
  \begin{center}
    \includegraphics[width=\textwidth]{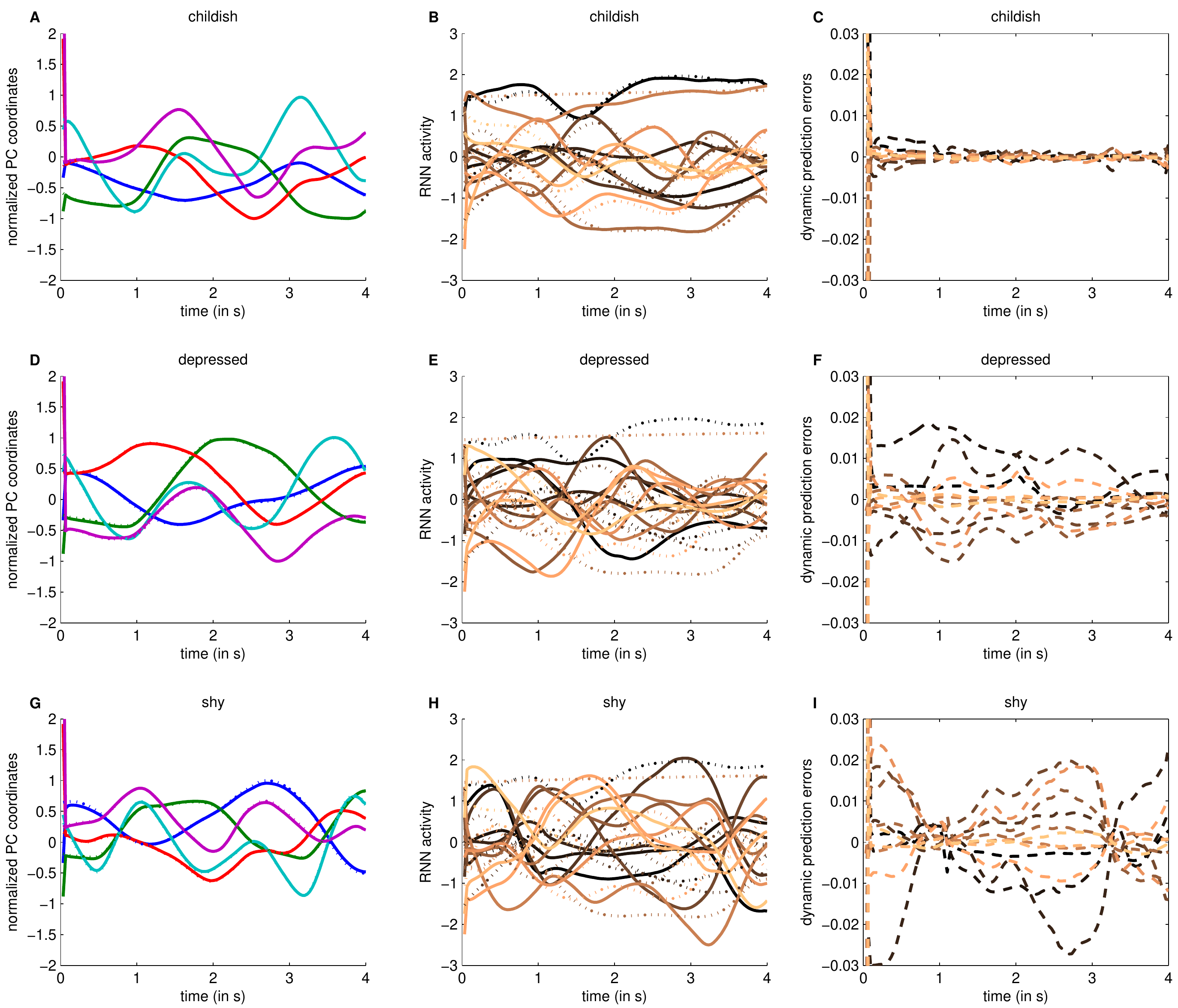}
  \end{center}
  \caption{
    Result of the three different walks recognized by one of the rRNNs (childish walk). (A,D,G) show the presented data (dotted lines) and the predicted sensory states (solid lines). (B,E,H) show the posterior hidden states (solid lines) and, for comparison, the hidden states of the corresponding gRNN when run autonomously from the initial states used during learning (dotted lines, cf. dotted lines in Fig. \ref{fig:randTraj}B). (C,F,I) show the dynamic prediction errors of the hidden states (Eq. \ref{eq:dynPredErr}, note that these prediction errors do not correspond to the difference between solid and dotted lines in the middle panels). The different rows of panels correspond to the different walks which were recognized (from top to bottom: childish, depressed and shy). Prediction errors were markedly lower, when the rRNN recognized the walk it was adapted for (C vs. F,I).
  }
  \label{fig:recognitionExmp}
\end{figure*}

In contrast, the rRNN based on this gRNN for the shy walk was robust against such differences in initial conditions. Even though we perturbed the rRNN initial states severely, the rRNN always switched rapidly to the appropriate dynamics which best described the sensory input of a shy walk. In other words, the prediction error updates of the hidden units forced the dynamics on a trajectory which predicted the observed walk. We depict a characteristic example of this quick response behaviour for the rRNN (childish walk) in Fig. \ref{fig:recognitionExmp} (A,B). After only one time step the rRNN accurately predicted all subsequent observations while hidden unit trajectories followed those typical for the learnt gRNN to a large extent. (Note that these results partially depend on an appropriate choice of the prior covariances, see Appendix C.) This means that the rRNN can represent the dynamic repertoire of the gRNN but, in addition, can rapidly switch to the specific dynamic regime that best explains the sensory input. 
\subsection{Discrimination of Dynamic Stimuli}
After learning, we have three different rRNNs, each of which has learnt to predict one of the three walking styles childish, depressed and shy. Here, we will show that the prediction error, on observations $\pe_\so = \vo -\bs{V}\vh$   or hidden states $\pe_\sh = \dot{\vh} - f(\vh)$ (Eq. \ref{eq:dynPredErr}), of all three rRNNs can be used to discriminate between the three different walks. In particular, we will show that the dynamic prediction errors $\pe_\sh$ are smallest for the rRNN that has learnt a specific walking style. This means that a potential readout mechanism can use the relative amplitudes of prediction error of the three rRNNs to decide which of the three walks is currently observed.

\begin{figure*}
  \begin{center}
    \includegraphics[width=\textwidth]{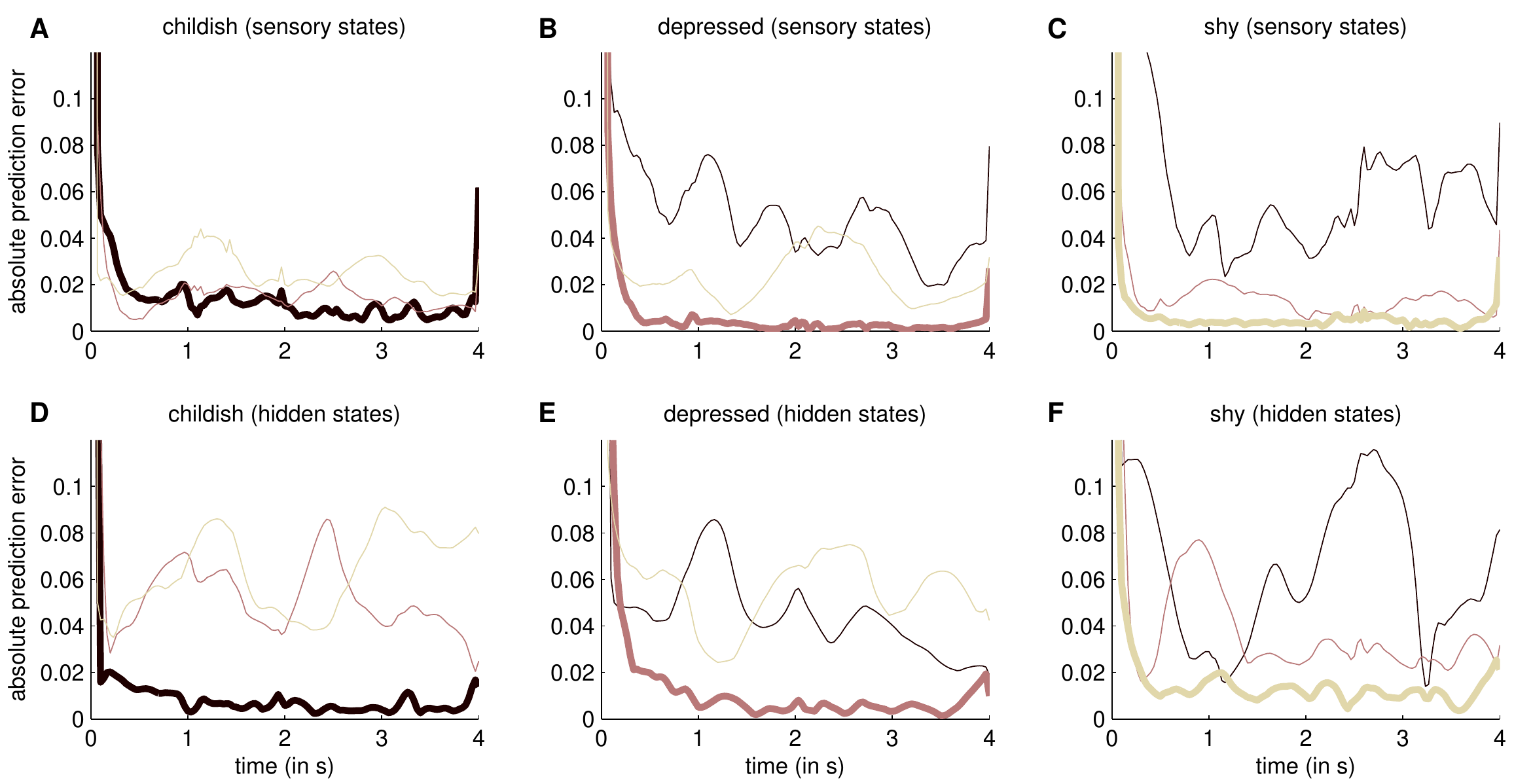}
  \end{center}
  \caption{
    Comparison of absolute prediction errors. Each panel shows summed (over state dimensions) absolute prediction errors of sensory (A-C) and hidden (D-F, Eq. \ref{eq:dynPredErr}) states of the three different rRNNs when data from one of the three different walks were observed. Each rRNN corresponds to one colour (childish: black, depressed: red, shy: yellow). Prediction errors of the rRNN, which has been learnt for the observed type of walk, are indicated by thick lines.
  }
  \label{fig:predErr}
\end{figure*}

Fig. \ref{fig:recognitionExmp} (D-I) shows the response of the rRNN which learnt the childish walk, but now given the depressed and shy walks as observations. Although this rRNN did not learn these walks it represented them well by exploiting alternative dynamics embedded in the twelve unit network. However, the rRNN frequently had to use prediction error on its hidden states to explain away the remaining mismatch between internal predictions and actual input. See Fig. \ref{fig:recognitionExmp} (C,F,I) for this relative increase in prediction error in response  to the non-learnt depressed and shy walks. This increase in prediction error when recognizing the two non-learnt walks is consistent over the three different rRNNs and may be used to discriminate dynamic stimuli as shown in Fig. \ref{fig:predErr} (D-F). For each of the three walks the prediction error was smallest for the rRNN which actually had learnt this specific walk, see also Table \ref{tab:predErr}. The prediction errors on observations showed this effect as well, although not as clearly (Fig. \ref{fig:predErr} A-C).
 \begin{table*}
  \caption{ Absolute prediction errors summed over time points and sensory, or hidden states, respectively. Top: sensory state prediction errors. Bottom: dynamic hidden state prediction errors. Each column presents results for each of the three different rRNNs on one of the three data sets childish, depressed and shy. rRNN (random): average accumulated prediction error obtained from 30 random rRNNs (values in parentheses show the minima). rRNN with lowest prediction error on each data set is indicated by bold font. Note that we excluded the first four out of 120 time points from these sums, because the initial transient period otherwise may distort the results.}
  \begin{center}
  \begin{tabular}{l r r r}
\multicolumn{4}{c}{sensory state prediction errors}\\
\hline
 & childish & depressed & shy\\
\hline
rRNN (childish) & $\mathbf{1.44}$ & $5.58$ & $6.81$\\
rRNN (depressed) & $1.69$ & $\mathbf{0.43}$ & $1.53$\\
rRNN (shy) & $2.89$ & $2.66$ & $\mathbf{0.56}$\\
rRNN (random) & $4.37 (2.37)$ & $4.30 (2.20)$ & $4.21 (2.57)$\\
\hline
\end{tabular}\vspace{1em}
\begin{tabular}{l r r r}
\multicolumn{4}{c}{hidden state prediction errors}\\
\hline
 & childish & depressed & shy\\
\hline
rRNN (childish) & $\mathbf{0.85}$ & $5.15$ & $7.38$\\
rRNN (depressed) & $5.95$ & $\mathbf{1.06}$ & $4.05$\\
rRNN (shy) & $7.39$ & $6.44$ & $\mathbf{1.48}$\\
rRNN (random) & $4.96 (3.04)$ & $4.86 (3.13)$ & $4.65 (3.03)$\\
\end{tabular}
  \end{center}
  \label{tab:predErr}
\end{table*}

We also investigated the effect of learning on the accumulated prediction errors by comparing the prediction errors of the learnt rRNNs with those of random rRNNs. We generated 30 random rRNNs by drawing random parameters $\bs{W}, \bs{V}$ and $\bs{k}$ while using the same connectivity constraints as for the gRNNs which were used for learning the walks. The accumulated prediction errors for the random rRNNs, thus, give an estimate for the total amount of prediction error expected in a random rRNN, i.e. without learning. As expected, the prediction errors of random rRNNs were always higher than those of the rRNNs with learnt parameters (see Table \ref{tab:predErr}). Furthermore, for non-learnt stimuli, the learnt rRNNs often produced larger prediction errors than random rRNNs. This  indicates that learning a specific walk restricts the dynamic repertoire of an rRNN. We conclude that the learning procedure resulted in rRNNs which were suited to discriminate the walks.

\begin{figure*}
  \begin{center}
    \includegraphics[width=\textwidth]{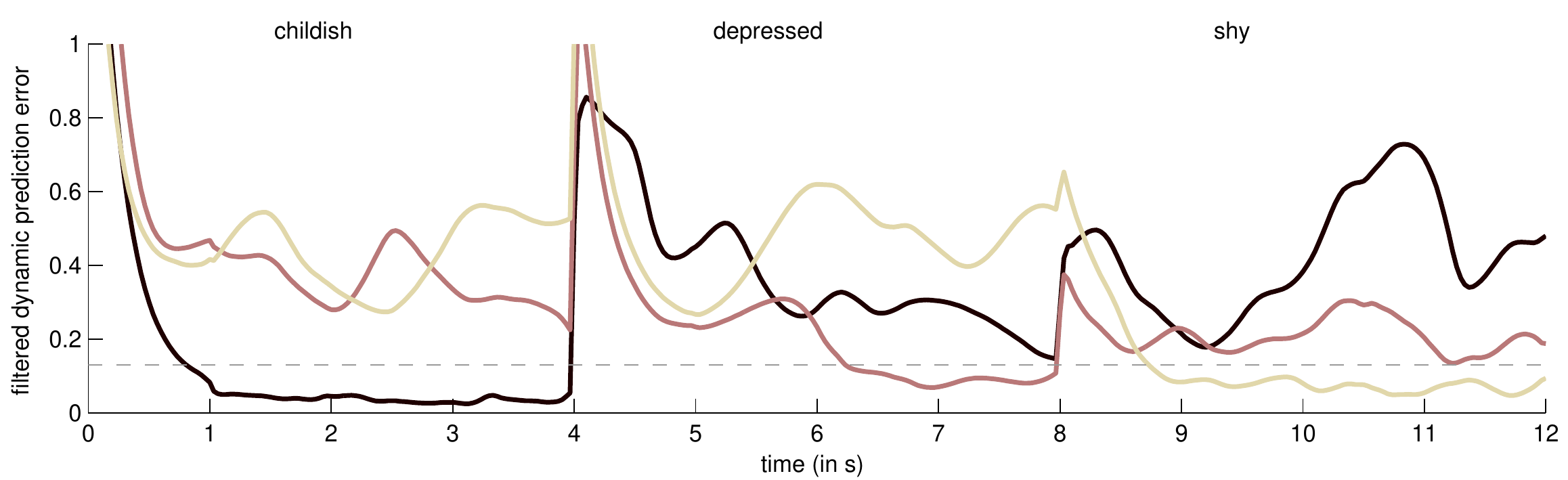}
  \end{center}
  \caption{
    Filtered Dynamic Prediction Errors. We concatenated sensory input  from all walks into a single sequence and inferred the hidden states for all three rRNNs. Shown are temporally smoothed, summed absolute prediction errors for the rRNNs learnt on the childish (black), depressed (red) and shy (yellow) walks. We also plotted an ad-hoc decision boundary (dashed, grey line) which could be used to select models with a low prediction error.
  }
  \label{fig:filtPE}
\end{figure*}

In an additional experiment we concatenated data from all three walks into a single sequence to simulate online recognition of three walks, see Fig. \ref{fig:filtPE}. The resulting saccade-like, abrupt transitions between walking styles led to a transient increase in prediction errors correctly signalling a discrepancy between predictions and actually observed kinematics. Furthermore, we implemented a simple readout mechanism for dynamic prediction errors using a filter which sums the absolute prediction errors over the last 30 time points and weights recent time points more strongly. This operation smoothes prediction errors temporally and stresses differences that stretch over a similar period as the filter size, see Fig. \ref{fig:filtPE}.  After each transition, the rRNNs reduced their smoothed prediction errors quickly until the rRNN with parameters learnt for the currently observed walk was the only one for which the magnitude of prediction errors fell below an ad-hoc threshold. This shows that the present approach can be successfully used to recognize 
a specific walk by choosing the model with the lowest prediction error, after some initial transient has died away.
\subsection{Robustness against Noise and Initial Conditions}
\label{sec:robust}
\begin{figure*}
  \begin{center}
    \includegraphics[width=\textwidth]{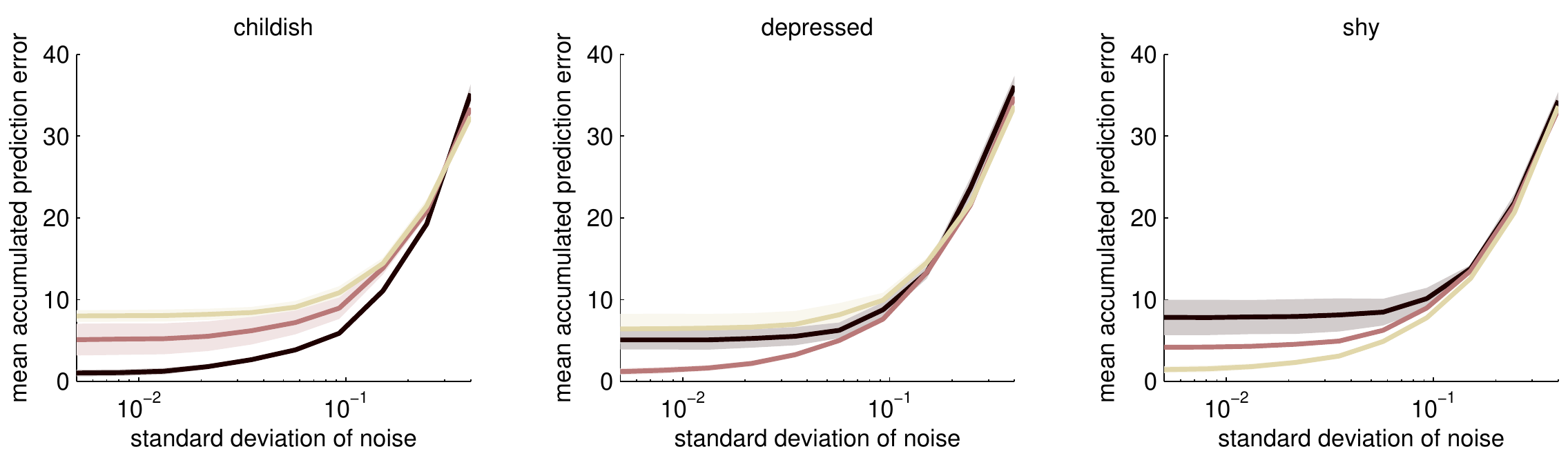}
  \end{center}
  \caption{Dependence of dynamic prediction errors on noise and initial conditions. Each panel shows average sums of absolute prediction errors for the three different rRNNs on one of the walks. The averages are over 12 randomly chosen initial states $\vh(0)^r$ and shading indicates the region around the mean of twice the standard deviation. The x-axis indicates the standard deviation of independent Gaussian noise added to the principal components of the walks on a log-scale. Note that the exponential increase of prediction errors with noise in the log-plot means that prediction errors depend approximately linear on the observation noise magnitude.
  }
  \label{fig:pEnoise}
\end{figure*}

Here, we demonstrate that the recognition scheme is robust to both noise and variations in initial conditions. We repeated the experiments above for increasing amounts of white observation noise and twelve different, randomly chosen sets of initial conditions, see Fig. \ref{fig:pEnoise}. We found that the overall magnitude of dynamic prediction errors is proportional to the amount of observation noise. This indicates that observation noise is explained away by prediction errors of both sensory and hidden units. Importantly, the discrimination ability of the three rRNNs is maintained up to moderate amounts of noise, i.e., prediction errors still contained sufficient information to discriminate the three walks. As expected, for large amounts of noise, the contribution from observation noise eventually masked the prediction error contributed by the difference in walks. Also note that the dynamic prediction errors of the learnt rRNNs on their learnt walks (bottom trajectories in the panels of Fig. \ref{fig:pEnoise}) had very low variability across initial conditions. This means that the rRNN, which was learnt for a specific walk and observes this walk as input, was much less dependent on its initial conditions than the rRNNs learnt on different walks. Yet, the variability of prediction error due to initial conditions within each rRNN was not large enough to influence the result of discrimination of the walks up to moderate amounts of noise. In other words, in our experiments accumulated prediction errors of the rRNN learnt for the current walk were always smaller than those of the other rRNNs (up to moderate amounts of noise), even when beneficial initial conditions for them led to better than average accumulated prediction errors.

\section{Discussion}
In this paper we have described the 'recognizing recurrent neural network' (rRNN), which is a recurrent neural network where each unit computes both predictions and prediction errors to recognize sensory input in a Bayes-optimal fashion. We derived the update equations of both sensory and hidden units by using an approximate Bayesian inference framework for nonlinear dynamical systems, i.e., dynamic expectation maximization \citep{Friston2008a}. The rRNN approach is motivated by two considerations: Firstly, in an rRNN, the additional prediction error messages, as compared to a conventional RNN, increase the complexity of the messages passed between units in the network. We propose that these computations may be better descriptions than conventional RNNs of what real neuronal ensembles implement. We base this hypothesis on recent findings and theoretical considerations which show that single neurons (and consequently neuronal ensembles) compute much more complex functions than previously thought \citep{Sidiropoulou2006,Spruston2008,Mel2008,Pissadaki2010,Debanne2011}. The general idea is that a single neuron may in principle compute complex, nonlinear and dynamic functions using its spatiotemporal voltage depolarisations and other dynamics like calcium fluctuations \citep{Mel2008}. Although it is yet unclear how the computation of predictions and prediction errors may map to cellular dynamics, intracellular dynamics may have in principle the computational complexity to perform Bayesian decoding of their synaptic input \citep{Deneve2008}. Secondly, the computation of predictions and prediction errors in an rRNN fits well with the recent reappraisal in cognitive neuroscience that the brain may use a predictive coding approach to perception and cognition \citep{Wassenhove2005,Summerfield2006,Bar2009,Friston2009}. The rRNN approach is a mathematical description of how this predictive coding scheme could be implemented for complex, multi-dimensional dynamic stimuli. 

To illustrate that rRNNs may be an interesting model for understanding the brain function of recognition and prediction for naturalistic stimuli, we showed that rRNNs can robustly recognize kinematics as observed with motion capture data. We found that the prediction error computed by an rRNN can be used to recognize and discriminate between different human walking movements in an online fashion. Furthermore, this recognition mechanism is robust against both noise on the observations and variations in the initial state of the rRNN. In other words, rRNNs may be used as functional models for human action observation studies, e.g. \citep{ Blake2007}. The rRNN approach unifies many important aspects of brain processing such as statistically optimal inference in highly variable and noisy environments, recurrent connections, online recognition of dynamics, and quick adaptation to sudden changes in the environment. Therefore, the rRNN approach may be a useful device to bridge the gap between behaviour-driven models of cognition and neurobiologically plausible models of neuronal ensembles. 

The idea to use autonomous RNNs as generative models is not entirely new. In previous work, we have used this approach in system identification where we explained neuroimaging data as generated by a network of cortical nodes, called 'Dynamic Causal Modelling' (DCM) \citep{Friston2003,Kiebel2006,Kiebel2009c}. Critically, the equations governing the dynamics of each node took the form of a rate model as in Eq. \ref{eq:gRNNdyn}. The difference to the present approach is that DCM uses specific, highly constrained connectivity schemes based on neural mass models and does not allow for errors in the hidden states. Similarly, we used the present approach \citep{Friston2008a} to model recognition of multi-scale dynamics \citep{Kiebel2008,Kiebel2009a} where the rRNN generalizes these previous contributions by using a more generic generative model (RNNs) and learning of natural stimuli. 

To our knowledge, the explicit use of (generic) recurrent neural networks as generative models for recognizing dynamic sensory input using online Bayesian inference has not been described before. Both techniques, Bayesian inference for dynamic stimuli and artificial recurrent neural networks have existed in parallel for many years now \citep{Jazwinski1970,Pearlmutter1989,Williams1989,Narendra1990}. We propose the combination of these two approaches in which RNNs act as dynamic models in a nonlinear, Bayesian filtering framework. Indeed, this idea has already been used implicitly in the field of machine learning and control. For example, \citet{Connor1994} used a related approach in the context of autoregressive models to remove outliers from sequences of discrete states which were represented by the hidden states of a RNN. Also, in dual extended Kalman filter methods for RNNs \citep{Wan2001} an extended Kalman filter is used to estimate RNN hidden states. However, these contributions focus on the usefulness of Bayesian filtering of RNN states to make conventional RNN learning more robust. Here we describe the idea that the combination of RNN equations and filter updates can themselves be interpreted as network equations which are better suited for recognizing dynamic stimuli. Therefore, the present approach also goes beyond previous suggestions of using RNNs as functions approximating the update equations of a nonlinear filter \citep{Parlos2001}, or its output \citep{Ting-HoLo1994}. We, thus, provide a novel perspective on the role of RNNs also in possible machine learning applications.

We motivated the present approach by considering the potential functional role of recurrently connected neuronal ensembles in cortical processing. This allowed us to address recognition of arbitrary nonlinear dynamics embedded in multidimensional, continuous stimuli - something that has not been reported with spiking neuron models of neuronal coding in recurrent networks \citep{Rao2004,Deneve2007,Wilson2009,Boerlin2011}. In contrast, the 'reservoir computing' approach \citep{Jaeger2001,Maass2002,Verstraeten2007} can recognize the class of stimuli we considered here. The reservoir computing approach has reinvigorated RNN research by establishing that very large networks (hundreds to thousands of units) combined with a simple readout function can be used to learn and recognize both dynamic and static stimuli. However, reservoir computing approaches typically do not adapt the dynamics within the network and rely on the chance probability that, among the many units, there exist some dynamical regimes which are appropriate generative models of the data. Here, we describe an alternative approach and speculate that small networks of 'smart' rRNN units may be sufficient to recognize dynamic stimuli.

Our use of RNNs as generative models of dynamic stimuli requires learning of their parameters ($\bs{W}, \bs{V}$ and $\bs{k}$ in eqs. \ref{eq:gRNNdyn}, \ref{eq:gRNNobs}). In particular, our results depend on learning the connections between hidden units in the recurrent network ($\bs{W}$). This type of learning has been proven to be difficult in the past \citep{Hammer2002}. While the learning procedure used here was capable of learning a sufficiently good dynamic representation of the present walks alternative learning approaches may have to be used to achieve similar performance on other data sets. For example, the different principal component coordinates of our walks had similar time scales (Fig. \ref{fig:learnResult}). More complex and longer movements may demand the use of hierarchical models and corresponding learning algorithms \citep{Hinton2006}.

While learning is an important issue with RNNs, we focused on providing a proof-of-principle that the rRNN approach can solve high-level problems such as discriminating visual dynamics in an online fashion. Importantly, our results appear to be robust against sub-optimally learnt generative RNNs. This can be seen in Fig. \ref{fig:learnResult}, which shows a residual difference between learnt trajectories and the input. In other words, we found that prediction error messages were sufficiently informative even though the sensory input observed by the rRNN deviated slightly from the internally predicted dynamics. We also found robustness of the discrimination against white observation noise, see Fig. \ref{fig:pEnoise}. It is an open question whether the rRNN approach is also robust against structured variations in human movements, e.g. as induced by a variation of a specific movement. We speculate that such variations require a different generative model, either where multiple movements are embedded in a single gRNN or where one uses a hierarchical gRNN similar to the approach used in \citet{Taylor2009}.

\appendix
\section{Bayesian Inversion of Dynamical Models}
In this appendix we give a high-level description of the D-step in Friston's dynamic expectation framework \citep{Friston2008a} which leads to Eq. \ref{eq:postmode} in the main paper.
\subsection{Generalised Coordinates}
A major component of Friston's approach to stochastic processes is the redefinition of the time-dependent variables in generalised coordinates of motion. For example, one replaces $\vh(t)$ with
\eq{
    \vhg(t) = \left[\vh(t)^\tr,\frac{\partial \vh(t)^\tr}{\partial t},\frac{\partial^2 \vh(t)^\tr}{\partial t^2}, \dots\right]^\tr
}
and obtains for the probabilistic form of Eq. \ref{eq:gRNNdyn} (dropping the dependence on $t$ for simplicity of writing)
\eq{\begin{split}
    \dot{\vh} &= \vh' = f(\vh) + \epsilon^\sh\\
    \frac{\partial^2 \vh}{\partial t^2} &= \vh'' =  \displaystyle \frac{\partial f}{\partial \vh} \vh' + \epsilon^{'\sh}\\
    \frac{\partial^3 \vh}{\partial t^3} &= \vh''' =  \displaystyle \frac{\partial f}{\partial \vh} \vh'' + \epsilon^{''\sh}\\\vdots
\end{split}}
Note that it is assumed that $f$ is locally linear around $\vh(t)$ and that differently from the usual stochastic process models dependencies between noise variables across time are allowed, i.e., it is assumed that the noise at two close points in time correlates and that the noise process $\epsilon^\sh(t)$ is differentiable sufficiently many times. In generalised coordinates of motion time-dependent variables encode not only a state at the current time, but additionally the future path of states. This is seen when we consider how the continuous representation here can be mapped onto a discrete sequence of $N$ future observations $\vo = [\so_1,\dots,\so_N]^\tr$ (for simplicity we here show only a single observed variable, i.e., $D=1$):
\eq{\label{eq:taylor}
    \so_i = \sum_{j=1}^n \frac{\sog_j(t)}{(j-1)!} (i-t)^{j-1}
}
where $n$ is the highest order of motion considered. We assume that $i=1,\dots,N$ are the times starting from $t$ at which the data have been sampled. This formula represents a Taylor series approximation making use of the derivatives in $\vog$. \citet{Friston2008a} have shown that the variance of the noise process quickly becomes very large for high order motions such that only a small number $n$ of generalised coordinates and data points need to be taken into account at any point in time. One also needs to translate discrete data samples into generalised coordinates of motion. This can be done using the inverse operation of Eq. \ref{eq:taylor}. Rewriting Eq. \ref{eq:taylor} in matrix form gives
\eq{\label{eq:gcconvert}\begin{split}
    \vo = \bs{E} \vog   \qquad    e_{ij} = \frac{(i-t)^{j-1}}{(j-1)!}\\  i \in \{1,\dots,N\},\quad j \in \{1,\dots,n\}.
\end{split}}
If $N=n$, $\bs{E}$ is invertible and one obtains $\vog(t) = \bs{E}^{-1}\vo$. The resulting $\vog(t)$ is then used to compute the likelihood of the data at $t$ and make inference over the hidden RNN states $\vhg(t)$ as described below.
\subsection{Dynamic Approximation of the Posterior Mode}
In generalised coordinates eqs. \ref{eq:gRNNdyn} and \ref{eq:gRNNobs} become
\eq{\label{eq:genmodel}
    \vog = \tilde{\bs{g}} + \tilde{\epsilon}_\so    \qquad    \bs{D}\vhg = \tilde{\bs{f}} + \tilde{\epsilon}_\sh
}
where $\bs{D}$ is a matrix differentiation operator which shifts coordinates upwards by one element, $\tilde{\bs{f}} = [\bs{f}^\tr,\bs{f}^{'\tr},\bs{f}^{''\tr},\dots]^\tr$ and $\tilde{\bs{g}} = [\bs{g}^\tr,\bs{g}^{'\tr},\bs{g}^{''\tr},\dots]^\tr$ are the predicted generalised states and observations, respectively, with $\bs{f}' = \frac{\partial f}{\partial \vh}\vh'$ and $\bs{g}' = \bs{V}\vh'$ (analogously for higher order terms $\bs{f}'',\dots$). Because $\tilde{\bs{g}}$ is linear here, one can write $\tilde{\bs{g}} = (\bs{I}\otimes \bs{V})\vhg$ where $\otimes$ denotes the Kronecker product and $\bs{I}\in \R^{n\times n}$. Based on these equations the log-likelihood of the observations $\vog(t)$ is defined as 
\eq{\begin{split}
    \mathcal{L}(t) &= \log p(\vog|\theta)\\
    &= \log \int p(\vog,\vhg|\theta) d\vhg\\
    &= \log \int p(\vog|\vhg,\theta) p(\vhg|\tilde{\bs{f}},\theta) d\vhg
\end{split}}
where $\theta$ is a placeholder for all parameters in the model. Notice that $p(\tilde{\bs{f}}(\vhg)|\Mo)$, where $\Mo$ represents previously seen data, has been approximated as $\delta(\tilde{\bs{f}}(\tilde{\mu}))$ - a Dirac delta function at $\tilde{\bs{f}}$ evaluated at the previous posterior mode $\tilde{\mu}$ (see below). This means that only the mode is propagated through the dynamics, but not its uncertainty. \citet{Friston2008a} then introduce a variational density $q(\vhg)$ (ignoring the density over parameters as learning is not our objective) and make use of Jensen's inequality to obtain
\eq{\label{eq:freee}
    \mathcal{L}(t) \geq \displaystyle \int q(\vhg) \log \frac{ p(\vog,\vhg|\theta)}{q(\vhg)} d\vhg = \mathcal{F}(q,t)
}
where $\mathcal{F}(q,t)$ is the free energy which is a lower bound on the log-likelihood. The aim is to find $q$ such that $\mathcal{L}(t) = \mathcal{F}(q,t)$. In other words, one maximises $\mathcal{F}(q,t)$ with respect to $q$. This is equivalent to minimising $KL[q(\vhg)|| p(\vhg|\vog,\theta)]$, the KL-divergence between variational density and true posterior, i.e., after optimisation $q$ is an approximation of the posterior density over RNN states. In particular, it can be shown \citep{Ghahramani2001,Friston2008a} that the $q$ maximising $\mathcal{F}(q,t)$ is equal to
\eq{\label{eq:qpost}\begin{split}
    q(\vhg) & = \frac{1}{Z} \exp(V(\vhg))\\
    & = \frac{1}{Z} \exp(\log p(\vog,\vhg|\theta))\\
    & = p(\vhg|\vog,\theta)
\end{split}}
where $V(\vhg)$ is called the variational energy. While this equation appears to be a trivial statement, the formulation of $q$ in this way lets us recognize \citep{Friston2008a} that $q$ also is the density defined by a set of stochastically moving particles at their stationary solution where the movement of a single particle is given by
\eq{
    \dot{\vzg} = \frac{\partial V(\vzg)}{\partial \vzg} + \Gamma(t) = \frac{\partial \log p(\vog,\vzg|\theta)}{\partial \vzg} + \Gamma(t)
}
and $\Gamma(t)$ is a random fluctuation. Using this relationship one can find $q$ using Monte Carlo simulation as we can compute the partial derivative of $\log p(\vog,\vzg|\theta)$. However, \citet{Friston2008a} simplified this further. In particular, a single particle in generalised coordinates with motion
\eq{\label{eq:mparticle}
    \dot{\vzg} = \kappa \frac{\partial V(\vzg)}{\partial \vzg} + \bs{D}\vzg
}
will converge to the mode $\tilde{\mu}$ of $V$, which is also the mode of the posterior, at a rate proportional to the constant $\kappa$ \citep{Friston2008a}. Given the mode $\tilde{\mu}$ \citet{Friston2008a} use a Laplace approximation for the posterior where $q \sim \mathcal{N}(\tilde{\mu},\tilde{\Sigma})$ is defined to be Gaussian and the covariance $\tilde{\Sigma}$ is found as
\eq{\label{eq:postcov}
    \tilde{\Sigma}^{-1} = \tilde{\Pi} = \left. - \frac{\partial^2 \log p(\vog,\vhg|\theta)}{\partial \vhg \partial \vhg}\right|_{\vhg = \tilde{\mu}}.
}
This is the inverse of the negative curvature of the posterior evaluated at the mode $\tilde{\mu}$. This completes the derivation of the approximate posterior over RNN states. 

Under the approximations made and given the linearity of $g$ one can identify the posterior $p(\vhg|\vog,\theta)$ as being Gaussian exploiting that $p(\vog|\vhg,\theta)$ and $p(\vhg|\tilde{\bs{f}},\theta)$ are Gaussian. In this case the Laplace approximation is exact. Nevertheless, we retained Friston's more general form which is also valid for nonlinear $g$. More importantly, this motivates the dynamic form of estimating the posterior mode in Eq. \ref{eq:mparticle} which allows us to extend the static result above to the dynamic case. In particular, note that all results above were obtained for only a single time point $t$. However, it can be shown \citep{Friston2008a} that the path integral of the free energy is maximised, if Eq. \ref{eq:qpost} holds for all $t$. Naively, this means that one has to integrate the motion of the particle in Eq. \ref{eq:mparticle} until it converges to $\tilde{\mu}(t)$ for each $t$. However, if the particle converges faster onto $\tilde{\mu}(t)$ than $\tilde{\mu}$ moves itself, a condition which can be ensured by choosing an appropriate rate constant $\kappa$, we will be able to track the motion of $\tilde{\mu}$ with a single particle and the dynamics given by Eq. \ref{eq:mparticle}. Intuitively, the representation in generalised coordinates of motion here helps to converge to a mode which better represents the data as it also takes the local motion (velocity, acceleration, etc.) of the mode into account.

For the purpose of this paper we ignored the approximated covariance and only concentrated on the posterior mode and the corresponding prediction errors. A summary of the resulting algorithm is shown in Table \ref{tab:alg}. We were able to ignore the covariance, because we assumed network parameters to be fixed during inversion. However, in the full DEM-framework these covariances are needed for the computation of parameter updates. 
\begin{table}
\caption{
Online algorithm for finding the approximate posterior of RNN states.}
{\tt
\begin{enumerate}
 \item[] initialise $\tilde{\mu}(0)$
 \item[] FOR $t$ = 1:$N$
 \item[] \begin{enumerate}
          \item[1)] compute predictions $\tilde{\bs{g}}(\tilde{\mu})$ and $\tilde{\bs{f}}(\tilde{\mu})$ from previous $\tilde{\mu}(t-\Delta t)$
          \item[2)] find $\vog$ based on $n$ data points closest to $t$ using Eq. \ref{eq:gcconvert}
          \item[3)] compute gradients of $V(\vhg)$ using predictions, $\bs{D}\tilde{\mu}(t-\Delta t)$ and $\vog$
          \item[4)] numerically integrate Eq. \ref{eq:mparticle} to get new $\tilde{\mu}(t)$
         \end{enumerate}
 \item[] END
\end{enumerate}
}
\label{tab:alg}
\end{table}

\section{Learning of RNN Parameters}
We want to adapt the RNN parameters $\bs{W},\bs{V},\bs{k}$ such that the observations generated by the RNN defined in eqs. \ref{eq:gRNNdyn},\ref{eq:gRNNobs} fit the data. We mainly follow the approach underlying dynamic causal modelling \citep{Friston2003, Kiebel2009c} which is detailed in \citep{Friston2002,Friston2002a}. This entails an iterative approximation of the parameter posterior based on a first-order Taylor expansion of an observation function $\vect(\Mo) = h(\theta)$ which represents the underlying dynamical system. Here, $\Mo\in\R^{N\times D}$ contains the observations at all $N$ time points and $\theta = [\vect(\bs{W})^\tr,\vect(\bs{V})^\tr,\bs{k}^\tr]^\tr$. The RNN states are enclosed in $h(\theta)$, because the dynamics is assumed to be noise free, i.e., deterministic. Both parameter likelihood and prior are assumed to be Gaussian so that the following gradients of the log-posterior $\mathcal{L} = \log p(\theta|\Mo) \propto \log p(\Mo|\theta)p(\theta)$ are obtained \citep[cf.][Eq. 17]{Friston2002}
\eq{\begin{split}
    \frac{\partial \mathcal{L}}{\partial \theta} &= \bs{J}^\tr \bs{C}_\so^{-1}\bs{r} + \bs{C}_\theta^{-1}(\mu_\theta - \hat{\theta}^{(i)})\\
    \frac{\partial^2 \mathcal{L}}{\partial \theta^2} &\approx \bs{J}^\tr\bs{C}_\so^{-1}\bs{J} + \bs{C}_\theta^{-1}.
\end{split}}
We use these in a numerical integration scheme for nonlinear dynamical systems to obtain an update of the parameters $d \theta$ based on the model $d\theta/dt = \partial \mathcal{L}/\partial \theta$ and $\hat{\theta}^{(i+1)}=\hat{\theta}^{(i)} + d\theta$. Here $\hat{\theta}^{(i)}$ is the maximum a posteriori estimate of the parameters in iteration $i$, $[\bs{J}]_{jk} = \partial [h(\hat{\theta}^{(i)})]_j/\partial \theta_k$ is the Jacobian of $h$ evaluated at $\hat{\theta}^{(i)}$, $\bs{C}_\so$ is the covariance of the observations and $\mu_\theta$ and $\bs{C}_\theta$ are the prior mean and covariance of the parameters, respectively. Finally, $\bs{r} = \vect(\Mo)- h(\hat{\theta}^{(i)})$ are the residuals of the data not explained by the predictions $h(\hat{\theta}^{(i)})$ which are equivalent to the observation prediction errors $\pe_\so$ described in the main text. In each iteration one obtains the predictions $h(\hat{\theta}^{(i)})$ by numerical integration of the RNN dynamics and the Jacobian $\bs{J}$ using numerical differentiation of $h(\hat{\theta}^{(i)})$.

In our experiments we divided learning into two phases - an initial phase in which we adapted parameters only on local chunks of the data and a final phase in which we used the complete data. We found that the first phase helped to find a better initialisation of $\hat{\theta}$ for the optimisation on the whole data set in the second phase of learning. In the first phase we split the data into seven overlapping, equal size chunks and ran two passes through all chunks where we ran only two iterations of the update procedure described above per chunk and pass. In the second phase we ran 25 iterations with a fast, approximate numerical integration scheme for $h$ and subsequently another 4 iterations with a slower, but more accurate scheme. While our choices for the number of chunks, passes and iterations led to good results, we expect that many other values may be chosen equivalently.

Embedded in each iteration there is also an expectation maximization (EM)-like update of hyperparameter $\lambda$ which determines the amount of noise on the observations $\bs{C}_\so = \mathrm{e}^\lambda \bs{I}$ during learning. We refer the reader to \citep{Friston2002,Friston2002a} for details. $\lambda$ was initialised as -32. The hyperprior for $\lambda$ was Gaussian $\lambda \sim \N(-\log(\bar{s}^2),1/8)$ where $\bar{s}^2$ is the average variance of the observation variables in the data. 

We initialised the parameters contained in $\theta^{(0)}$ as follows. The elements of $\bs{k}^{(0)}$ were chosen uniformly in the interval $[1/8,3/8]$. Randomly chosen 2/3 of all elements in $\bs{W}^{(0)}$ were fixed at 0, the remaining were drawn randomly from a standard normal distribution. Furthermore, following \citep{Jaeger2007}, we scaled the resulting matrix by $\bs{W} = 1/(0.95\delta)\bs{W}^{(0)}$ to bring the initial RNN dynamics to a useful dynamical regime. $\delta$ here is the largest absolute eigenvalue of the matrix $[\bar{k}\bs{W}^{(0)}-(1-\bar{k}a)\bs{I}]$ where $a$ is the leakage (cf. Eq. \ref{eq:RNNdyn}) and $\bar{k}=1/4$ is the expected value of $k_i$ for any $i$. The initial states $\sh_j(0)^l$ were chosen uniformly in $[-2,2]$ for all $j$. We then found $\bs{V}^{(0)}$ as the solution to the underdetermined system of equations $\vo(0) = \bs{V}^{(0)}\vh(0)^l$ using Matlab's backslash operator, i.e., we found the least squares solution for $\bs{V}^{(0)}$ with most elements of $\bs{V}^{(0)}$ equal to zero. A randomly chosen subset of these zero elements were also fixed during learning. The number of fixed elements was 1/3 of the total number of elements. 

In the initial learning phase we set the mean of the parameter prior to the described initialisation of the parameters $\mu_\theta = \theta^{(0)}$. In the subsequent learning phase we set $\mu_\theta$ to the result of the 1st phase. The covariances of the prior parameter distribution were chosen to be diagonal, but also differed in the two phases of learning. In the initial phase we set the variances associated with the elements of $\bs{W}$ to $1.6\cdot 10^5$ while we set the variances for $\bs{V}$ and $\bs{k}$ to $0.018$ and $0.135$, respectively. This enforced particularly the adaptation of the dynamical parameters. For learning on the full data set we chose these variances to be $7.389$, $1$ and $1$ for $\bs{W}$, $\bs{V}$ and $\bs{k}$, respectively.
\section{Prior Covariances}
For the rRNN the prior covariances, $\tilde{\Sigma}_\so$ and $\tilde{\Sigma}_\sh$, modulate the size of updates of the posterior (cf. eqs. \ref{eq:Vgrado} and \ref{eq:Vgradh}) and influence the result of the Bayesian inversion. Intuitively, for large prior (co-)variances, i.e., a large amount of a priori expected noise, smaller updates are made and larger prediction errors are tolerated. The amount of noise here has to be seen in comparison to the variance of the unperturbed states of the units in the gRNN. For the sensory states this corresponds to the variance of the movement data. The standard deviation of the sensory states across all walks averaged over the five input dimensions was 0.38 while the standard deviation of the corresponding changes in hidden states averaged over the 12 hidden units was 0.04. In our simulations we assumed isotropic prior noise and correspondingly chose covariances of the form $\Sigma_\so = \sigma_\so^2\bs{I}$ \footnote{To see how these standard covariance matrices are translated into generalised coordinates we refer the reader to \citep[p. 860]{Friston2008a}.} where $\bs{I}$ is the identity matrix and $\sigma_\so$ is our choice of standard deviation. We chose $\sigma_\so=0.3$ and $\sigma_\sh=0.1$ for sensory and hidden states, respectively. This means that we tolerated only relatively small prediction errors on sensory states while allowing for relatively larger prediction errors on changes of hidden states. This choice implements the natural prior belief that the variability of walk observations is mainly determined by the variability of the underlying dynamics.
\bibliographystyle{abbrvnat}
\bibliography{rRNNpapers}

\end{document}